\documentclass{iau}
\usepackage{graphicx}
\graphicspath{{./Figures/}}
\usepackage{caption}

\usepackage{natbib}
\bibpunct{(}{)}{;}{a}{}{,}    %
\usepackage{soul}
\usepackage[dvipsnames]{xcolor}
\usepackage[colorlinks=true,linkcolor=MidnightBlue,citecolor=MidnightBlue,urlcolor=MidnightBlue]{hyperref}

\newcommand{\RH}{R_{\rm H}}
\newcommand{\Tsurv}{\ensuremath{T_{\mathrm{surv}}}}
\newcommand{\Ms}{\ensuremath{\mathrm{M}_\odot}}

\newcommand{\perrat}[1]{\ensuremath{\nu_{#1}}}
\renewcommand{\epsilon}{\varepsilon}

\renewcommand{\H}{\mathcal{H}}
\newcommand{\K}{\mathcal{K}}
\newcommand{\deriv}[2]{\frac{\mathrm{d} #1}{\mathrm{d} #2}}

\newcommand{\theres}{\ensuremath{\theta_\mathrm{res}}}

\newcommand{\genperrat}{\ensuremath{\nu}}

\newcommand{\rescoefpq}{R_{pq}}

\newcommand{\plsep}{\delta}
\newcommand{\resloc}{\ensuremath{\eta}}
\newcommand{\freqpq}{\omega_{pq}}

\newcommand{\Mfac}{\ensuremath{M}}

\newcommand{\reswidth}[1]{\ensuremath{\Delta #1}}
\newcommand{\reseta}{\reswidth{\resloc_{pq}} }

\newcommand{\numfacres}{6.55}

\newcommand{\density}{\ensuremath{\rho}}
\newcommand{\denstot}{\ensuremath{\density_{\mathrm{tot}}}}

\newcommand{\resind}{\ensuremath{k}}

\newcommand{\plsepov}{\ensuremath{\plsep_{\mathrm{ov}}}}

\newcommand{\ovind}{\ensuremath{k_{\mathrm{ov}}}}

\newcommand{\diffcoef}[1]{\ensuremath{D_{#1}}}
\newcommand{\diffcoefeff}{\diffcoef{\mathrm{eff}}}

\doi{10.1017/xxxxx}

\volume{382}
\jname{Complex Planetary Systems II}
\editors{A. Lemaitre, A.-S. Libert, eds.}

\title[Stability of compact planetary systems]{Long-term stability and dynamical spacing of compact planetary systems}

\author[A. C. Petit]{Antoine C. Petit}
\affiliation{Université Côte d'Azur, Observatoire de la Côte d'Azur, CNRS, Laboratoire Lagrange, France\\
email: {\tt antoine.petit@oca.eu}}

\begin{document}

\maketitle

\begin{abstract}
  Exoplanet detection surveys revealed the existence of numerous multi-planetary systems packed close to their stability limit. 
  In this proceeding, we review the mechanism driving the instability of compact systems, originally published in \citep{Petit2020a}. 
  Compact systems dynamics are dominated by the interactions between resonances involving triplets of planets. 
  The complex network of three-planet mean motion resonances drives a slow chaotic semi-major axes diffusion, leading to a fast and destructive scattering phase. 
  This model reproduces quantitatively the instability timescale found numerically. 
  We can observe signpost of this process on exoplanet systems architecture.
  The critical spacing ensuring stability scales as the planet-to star mass ratio to the power 1/4.
  It explains why the Hill radius is not an adapted measure of dynamical compactness of exoplanet systems, particularly for terrestrial planets.
  We also provide some insight on the theoretical tools developped in the original work and how they can be of interest in other problems.
\end{abstract}

\firstsection
\section{Introduction}

Exoplanets discoveries have shown that one of the most common type of planetary system is composed of multiple Super-Earths orbiting very close to their host star \citep[\emph{e.g.}][]{Fabrycky2014}.
The orbits of these planets are almost circular and coplanar \citep{Johansen2012,Xie2016,He2020} and the large majority of them are not in resonance \citep{Izidoro2017}.
Moreover these systems have a strong intra-system uniformity: planet radii and spacings between the planets are correlated with their neighbours \citep{Weiss2018,Murchikova2020}.
The spacing between the planets is constrained by the dynamical stability as there is a lower limit to the distance between two orbits at around 10 Hill radii \citep{Pu2015}.

Understanding the origin of this stability boundary has been the topic of many research works.
Indeed, the mutual Hill radius $ \RH = (a_1+a_2)/2((m_1+m_2)/(3m_\star))^{1/3} $, where $a_k$, $m_k$ are respectively the planet  semi-major axis and mass, is the natural length unit in the planetary three-body problem, as it measures the distance where the planet-planet interactions start to dominate over the stellar gravity.
In this problem, the Hill stability gives an absolute boundary \citep{Marchal1982} preventing from planet close encounters and collisions (but not from ejections or star-planet collsions).
This boundary, which can be expressed \citep{Petit2018} as a condition on the total Angular Momentum Deficit \citep{Laskar1997,Laskar2017} can be simplified to a minimum spacing between circular and coplanar orbits \citep{Gladman1993} such that $a_2-a_1>2\sqrt{3}\RH$ guarantees the longterm stability of the system.

This sharp boundary has no equivalent in systems with more than three planets.
\cite{Chambers1996} found numerically that multiplanetary systems experience a long quiescent phase where the system dynamics are almost secular before a very rapid transition to collisional dynamics.
The instability timescale also scales exponentially with the spacing between the planets.
While \cite{Chambers1996} proposed to scale the spacing with the mutual Hill radius of the planets, they also showed that a scaling as $(m_{\rm pl}/m_\star)^{1/4}$ is more accurate over a large mass ratio regime.
Many works further confirmed that pionneering result \citep[\emph{e.g.}][see \citealp{Petit2020a} for a comprehensive review]{Faber2007,Smith2009,Obertas2017} but most authors sticked with the Hill radius as a way to rescale the problem.

In \cite{Petit2020a}, we proposed the first quantitative analytical model of the instability mechanism in compact multiplanetary systems.
Building upon the work of \cite{Quillen2011}, we derived a criterion for the overlap of a mean motion resonance (MMR) network involving three planets.
We also showed that the effective diffusion of the semi-major axis due to this resonant network has a timescale consistent with numerical results.
Moreover, the critical spacing for stability between the planets scales as $(m_{\rm pl}/m_\star)^{1/4}$.

In this proceeding, we come back on the results from \cite{Petit2020a} and discuss how the concepts developped in this study can be used beyond this case.
In Sect. \ref{sec:review}, we briefly recall the main findings that work.
We show in Sect. \ref{sec:spacing} that the dynamical spacing measurement coming from our model provide a better descritpion than the Hill spacing to understand the instability limit of exoplanet systems.
In particular, we show that exoplanets systems are clustered much closer to the instability than if the spacing is measured in Hill radii.
In Sect. \ref{sec:optdepth}, we discuss in detail why the idea of resonant optical depth is so powerful and the conditions where such a tool can be useful, in particular to compute analytically diffusion timescales.

\section{The path to instability in compact systems}
\label{sec:review}

This section summarizes the results from \cite{Petit2020a}, we refer to it for a more detailed description.
We study a system of three planets of masses $m_1,m_2, m_3$ orbiting a star of mass $m_\star$.
The orbits are assumed to be circular and coplanar (see Sect. \ref{sec:ecc} for a discussion on generalization to the non-idealized case).
The system coordinates are the Delaunay coordinates $(\Lambda_i,\lambda_i)$ where $\Lambda_i=m_i\sqrt{\mu a_i}$, $a_i$ is the planet semi-major axis and $\mu = \mathcal{G} m_\star$.
We consider systems out of resonant chains, whose instabilities are studied in \citep{Pichierri2020}.
While most numerical studies on this problem focus on the equal spacing case, we do not make this assumption, simply assuming that the spacing is small with respect to the semi-major axes.

\subsection{Instability mechanism}

The key features the tightly packed system instability present can be summarized as
\begin{enumerate}[a.]
  \item The survival time $\Tsurv$ has an exponential dependency on the orbital spacing, measured in units of $(m_{\rm pl}/m_\star)^{1/4}$. 
  The fit is valid over 6 to 8 orders of magnitude for $\Tsurv$ between 100 and $10^{10}$ orbits.
  \item Instabilities occur for spacings larger than the Hill stability criterion, it is an intrinsically multi-planet phenomenon.
  While there is a small quantitative change going from three planet to five planet systems, there is no dependency beyond that number \citep{Chambers1996}.
  Three planets are necessary but also sufficient to reproduce the effect.
  \item The survival time distribution suggests that the evolution is driven by a diffusion process \citep{Hussain2020}. The dips close to first-order two-planet MMRs indicate that these latter play a fundamental role in enabling the orbit crossing.
  \item Unstable systems spend most of their lifetime in a quiescent phase where they only experience weak diffusion of their semi-major axes but no AMD evolution. Once a system encounters a two-planet first-order MMR, a rapid chaotic phase leads to orbit crossing.
\end{enumerate}

\begin{figure}
\begin{minipage}[c]{0.47\linewidth}
  \begin{center}
	\includegraphics[width=1.1\linewidth]{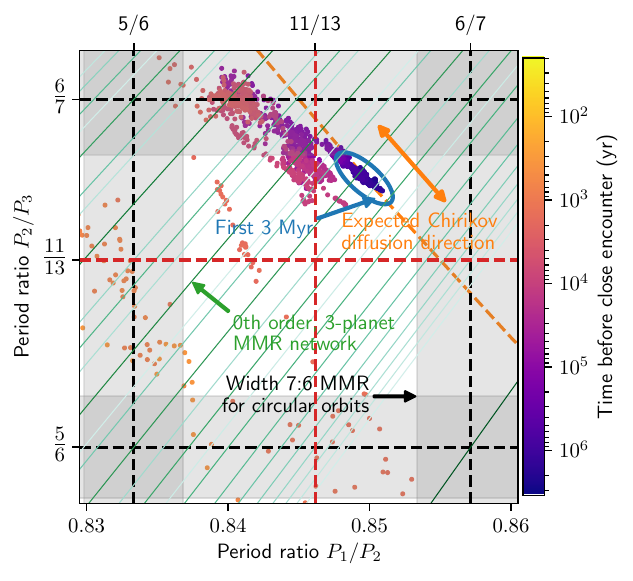}
\end{center}
        \captionof{figure}{Evolution in the period ratios plane. The points are colour-coded according to their time before the close encounter.
        We note that the system spends almost all of its time very close to its starting location.
        Green oblique lines correspond the loci of the zeroth-order three-planet MMRs.
        The first-order two-planet  MMRs, 7:6 and 6:5, are plotted in grey.
        The second-order resonance 13:11 is plotted in red. Adapted from \cite{Petit2020a}. \label{fig:phen} }
      \end{minipage}\hfill%
  \begin{minipage}[c]{0.47\linewidth}
  \begin{center}
    \includegraphics[width=1.1\linewidth]{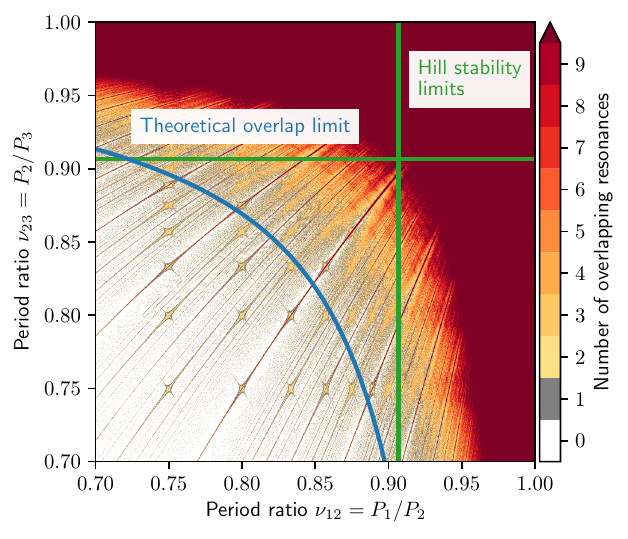}
    \captionof{figure}{Number of resonances covering the period ratios space for equal-mass planets, with masses $10^{-5}$ \Ms.
At first order, the number of resonances can be compared to the resonance optical depth $\denstot$.
    We plot the two-planet circular Hill-stability limits \citep{Petit2018} for both planet pairs in green and the predicted overlap limit for the three-planet MMRs (Eq. \ref{eq:plsep-ov}) in blue. Adapted from \cite{Petit2020a}.
\label{fig:overlap}}
  \end{center}
\end{minipage}
\end{figure}

We illustrate the mechanism leading to instability on a typical numerical simulation.
In this case, the planet masses are equal to $m_{\rm p }= 10^{-5}\ \Ms$ and the period ratio is initially close to  $1.175$.
We reproduce in Fig. \ref{fig:phen}, the Figure 3 of \cite{Petit2020a}.
This Figure shows the evolution of the period ratios of both pair of planets $P_1/P_2$ and $P_2/P_3$ during the system lifetime.
Planets start outside of the main neighbouring MMRs (the 6:5 and 7:6) and remain close to the initial position for 3 Myr.
They diffuse towards the 7:6 MMR between the second and third planet perpendicularly to zeroth-order three planet MMRs (see below) as expected due to \cite{Chirikov1979} diffusion.
Once the 7:6 MMR is reached, a close encounter is almost immediate.

\subsection{Three-planet mean motion resonances overlap}

Based on \cite{Quillen2011}, we postulate that the slow diffusion is driven by the overlap of three-planet zeroth-order MMR.
A zeroth-order three-planet MMR can be described by two integers $p$ and $q$. The resonance equation is
\begin{equation}
pn_1-(p+q)n_2+qn_3 = 0 \mathrm{\quad or\, equivalently \quad}\perrat{23} = 1-\frac{p}{q}(\perrat{12}^{-1}-1),
\label{eq:03pMMR}
\end{equation}
where $\perrat{ij} = P_i/P_j = n_j/n_i$, $p$ and $q$ are integers and, $n_j$ is the planet mean motion. The resonance loci in the period ratio plane $(\perrat{12},\perrat{23})$ correspond to hyperbola going through the point (1,1). and the full network  is dense in that plane. 

The resonance loci do not depend on the MMR index $p+q$, but only on the ratio of the integers $p/q$.
An adapted set of coordinates to describe the period ratio plane can be defined to take advantage of this property.
We define the resonance locator $\resloc$ and the generalised period ratio $\nu$
\begin{equation}
\resloc = \frac{1-\perrat{23}}{\perrat{12}^{-1}-\perrat{23}}  \mathrm{\quad and \quad}  \genperrat = \frac{(1-\perrat{12})(1-\perrat{23})}{1-\perrat{12}\perrat{23}}.
  \end{equation}
The resonant locator $\resloc$ is a smooth function of the period ratio and $\resloc = p/(p+q)$ on the resonance defined by $(p,q)$. 
Constant $\genperrat$ lines are hyperbola along which the resonance strength is roughly comparable for a given $p+q$.
Inspired by the generalized period ratio $\nu$, we can also define a generalized spacing between the planets that the takes the form
\begin{equation}
	\plsep = \frac{\plsep_{12}\plsep_{23}}{\plsep_{12}+\plsep_{23}} = \left(\frac{1}{1-\alpha_{12}}+\frac{1}{1-\alpha_{23}}\right)^{-1} \simeq \frac{2}{3}\genperrat,
	\label{eq:plsep-def}
\end{equation}
where $\plsep_{ij} = 1-\alpha_{ij}$ and $\alpha_{ij}=a_i/a_j$.

When considering a single resonance defined by integers $(p,q)$, we can show that there exist a canonical transformation from the Delaunay coordinates to a set of coordinates $(\Theta,\Gamma,\Upsilon,\theres,\theta_\Gamma,\theta_\Upsilon)$ such that $\Gamma,\Upsilon$ are first integral of the motion after averaging over the fast angles to the second order in mass \citep{Petit2020a}.
In these coordinates, the motion is described by a classical pendulum Hamiltonian
  $\H_\mathrm{res} = -\K_2/2(\Theta-\Theta)^2 +\epsilon^2R_{pq} \cos \theres$
where $\Theta_0$ is the value of the action $\Theta$ at the resonance locus for given $\Gamma$ and $\Upsilon$, $\epsilon$ is a factor proportional to the planet-to-star mass ratio to remind the order of each term,
$\K_2$ is the second-order expansion of the Keplerian part around the resonance center and
$\epsilon^2\rescoefpq$ is the factor in front of the resonant term.
For these two last terms, we refer to Eqs. (42,47,49) of \cite{Petit2020a}.
The width of the resonance and the small amplitude libration are straightforward to compute, we have respectively $\reswidth{\Theta} = 2\epsilon\sqrt{2\rescoefpq/\K_2}$ and $\freqpq = \epsilon\sqrt{\K_2R_{pq}}$.

The resonances have a clearer geometrical interpretation in the period ratio space, particularly when one needs to compare them.
We compute the width of the resonances perpendicularly to the network, that is, the width in terms of the variable $\resloc$
\begin{equation}
\reseta = \deriv{\eta}{\Theta}\reswidth{\Theta} =  \numfacres\epsilon\Mfac \frac{(\resloc(1-\resloc))^{3/2}}{\plsep^{2}}e^{-(p+q)\plsep},
\label{eq:reswidtheta}
\end{equation}
where 
  \begin{equation}
    \epsilon \Mfac = \frac{\sqrt{m_1m_3+m_2m_3\eta^2\alpha_{12}^{-2}+m_1m_2\alpha_{23}^{2}(1-\eta)^2}}{m_\star}\label{eq:Mfac}
    \end{equation}
describes the planet mass dependency.
Importantly, the width $\reseta$ only depends on the sum $p+q$ such that the width is a smooth function for the subnetworks having the same resonant index $p+q$.

In order to find the region where the resonances overlap, we rely on the concept of resonance network optical depth or filling factor developped by \cite{Quillen2011} and applied in the two-planet case by \cite{Hadden2018}.
The idea is to consider the local proportion of the space that is occupied by the MMRs. 
If this proportion is larger than 1, it is a good approximation to consider that MMRs fully overlap and a diffusion of the actions is possible.
We detail more in Sec.~\ref{sec:optdepth} the conditions where such concept is useful.
Resonances in subnetworks with constant index $p+q$ have comparable size and a constant spacing in terms of $\resloc$, such that the expression of the optical depth of the subnetwork is easy computed $\rho_{p+q}=(p+q)\Delta\resloc$.
The total optical depth of the network is bounded by the sum over $k=p+q$ of the optical depth of all subnetworks $\denstot \leq \sum_k \density_k$ as some resonances are counted with multiplicity.
Yet, the exponential decay in $k$ allow us to replace the inequality by an equality.
Substituting the sum by an integral, the optical depth is
\begin{equation}
	\denstot = \numfacres\epsilon\Mfac \frac{(\resloc(1-\resloc))^{3/2}}{\plsep^{2}} \int_{0}^{+\infty}\resind e^{-\resind\plsep}\mathrm{d} \resind	=\numfacres\epsilon\Mfac\frac{(\resloc(1-\resloc))^{3/2}}{\plsep^{4}}.
\label{eq:denstot}
\end{equation}
As \cite{Quillen2011}, we find that the filling factor depends linearly on the mass ratio and scales as $\plsep^{-4}$.
We confirm that the natural spacing rescaling for the problem is not the Hill radius, which scales as $\epsilon^{1/3}$, but rather a dependency on $\epsilon^{1/4}$.\
We can define a critical spacing value $\plsepov$ such that the zeroth-order three-planet MMR network fills the entire space. Taking $\denstot=1$ and solving for $\plsep$, one obtains
\begin{equation}
\plsepov = 1.59(\epsilon M)^{1/4}(\resloc(1-\resloc))^{3/8}.
	\label{eq:plsep-ov}
\end{equation}
Here, $\plsepov$ is a function of the masses and $\resloc$. 
In the case of equal mass and spacing systems, Eq. (\ref{eq:plsep-ov}) becomes $\plsep_{\mathrm{ov,eq}} = 1.0\left(m_p/m_0\right)^{1/4}$\footnote{There is a typo in \citep{Petit2020a} in Eq. (61) noticed by D. Tamayo (private communication).}.

We plot in Figure \ref{fig:overlap}, the number of resonances that overlap at a given point in the plane $(\perrat{12},\perrat{23})$ for three equal-mass planets as well as the two-planet Hill stability limits and the MMR overlap criterion derived in Eq. (\ref{eq:denstot}).
The number of resonances is to first order a proxy for the filling factor \denstot.
We see in Figure \ref{fig:overlap} that the region where the overlap of the three-planet MMR network takes place extends well beyond the Hill-stability limits, particularly for comparable spacings between the two neighbouring planet pairs.

\subsection{Diffusion model and instability timescale}
\begin{figure}
  \begin{center}
  \includegraphics[width=0.6\linewidth]{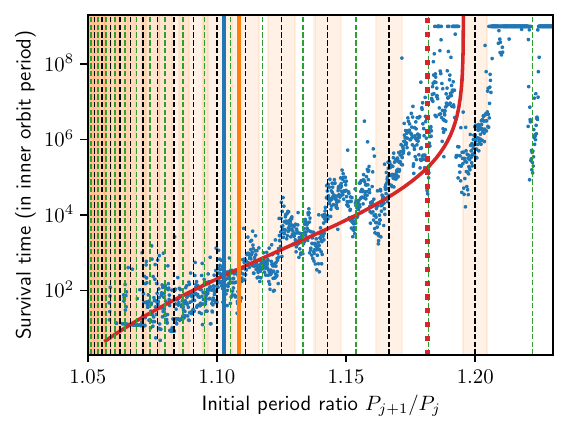}
\caption{Survival time in numerical simulations for a equal spacing and $10^{-5}$  \Ms{} mass three-planet system as a function of the initial period ratio. The red curve corresponds to the survival time estimate (eq. \ref{eq:Tsurv_estimate}), the blue vertical line to the Hill-stability limit \citep{Petit2018}, the orange vertical line to the two-planet MMR overlap criterion \citep{Wisdom1980}. The dashed black (resp. green) lines are the two-planet first(resp. second)-order MMR. The light orange rectangles show an estimate of the width of the two-planet MMR \citep{Petit2017}. We refer to \cite{Petit2020a} (from which this firgure is adapted) for the description of the numerical setup.\label{fig:EMS1e5}}
\end{center}
\end{figure}

When resonances overlap, \cite{Chirikov1979} predicts that the actions diffuse perpendically to the resonance hyperplane with a diffusion rate that depends on the resonance width and the period of the resonance such that the diffusion depends expoentially in the resonance index.
As the resonance index increases, the associated diffusion rate vanishes, such that in the limit where the diffusion is dominated by smaller and smaller resonances, the timescale effectively tends to infinity.
However, for $\plsep<\plsepov$, not all the resonances are necessary to cover the phase space.
We therefore only need to consider the largest ones to compute the survival time. 

The diffusion in the $\resloc$ direction is quite representative of the general Chirikov direction while giving a good parametrization of the resonant network.
Since the diffusion rate depends on the index of the resonance, the diffusion is not homogeneous and the diffusion of $\resloc$ follows the equation
\begin{equation}
	\deriv{\resloc}{t} = \sqrt{D(\resloc)}b(t),
	\label{eq:mod-Langevin}
\end{equation}
where $D(\resloc)$ represent the diffusion coefficient due to the lowest index resonance at this location.
\cite{Morbidelli1997} give a solution to Eq. (\ref{eq:mod-Langevin}) providing that we can describe the function $D(\resloc)$.
However, in our case, this function is highly variable and we instead use a probabilistic approach.
Instead of following the random walk by looking at which exact resonances are crossed until the first close encounter, we obtain a mean survival time by averaging over all the possible combinations of resonances that can contribute to the diffusion process (\emph{i.e} the MMR with index lower than $\ovind$).
In particular, we can define an effective diffusion coefficient, taking into account the contribution of all the resonances necessary to locally cover the phase space
 \begin{equation}
	\diffcoefeff = \left(\int_0^{\ovind} \frac{\density_k}{\sqrt{\diffcoef{k}}}\mathrm{d} k \right)^{-2}= 
	\left(\int_0^{\ovind} \frac{k}{\sqrt{\omega_k}}\mathrm{d} k \right)^{-2}.\label{eq:diffcoef1}
 \end{equation}
While the chaotic diffusion will be much more heterogenous (with fast and slow phases), on average the system will behave the same way if it where subject to a constant diffusion process of parameter $\diffcoefeff$.
Taking into account the distance that the system needs to travel until it reaches a two-planet MMR that destabilizes it for good, we obtain a mean survival time for compact systems where three-planet zeroth-order MMR overlap
\begin{equation}
	\left\langle \log\frac{\Tsurv}{P_1}\right\rangle \simeq  -\log\left(26.2\epsilon\Mfac\sqrt{\resloc(1-\resloc)}\frac{1-(\plsep/\plsepov)^4}{(\plsep/\plsepov)^6}\right)
	+\sqrt{\left|\ln 1-\left(\frac{\plsep}{\plsepov}\right)^4\right|}.
  \label{eq:Tsurv_estimate}
\end{equation}
This expression can be decomposed into a prefactor that mainly depends on the planet-to-star-mass ratio and a function that only depends on $\plsep/\plsepov$.
At first glance, this expression is not linear in $\plsep$, however the linear approximation is correct in the regime of interest, that is, for Hill-stable planet pairs not too close to the overlap limit.
As \cite{Quillen2011}, we predict that beyond the overlap limit, the survival time is likely much larger since the Chirikov diffusion is not an efficient process on its own.

To compare with previous numerical studies, we compute the estimated survival time for equal-mass and equally spaced planets $\log_{10}(\Tsurv/P_1) \simeq -\log_{10}\epsilon_p-6.51+3.56\Delta/\epsilon^{1/4}_p$, where $\epsilon_p=m_p/m_\star$.
The slope coefficient 3.56 is very close to values obtained in previous works and as already discussed the spacing is rescaled by $\epsilon^{1/4}_p$.
The prefactor proportional to $1/\epsilon_p$ is consistent with numerical simulations and the numerical constant is very close to the one obtained by \cite{Faber2007}.
We show on Fig. \ref{fig:EMS1e5} a direct comparison to numerical simulations.
We see that the analytical expression captures most of the trends observed, including the change of behaviour where the network stops overlapping.
A more in-depth comparison to numerical results is done in \cite{Petit2020a}, from planet mass of $10^{-7}$ \Ms to $10^{-3}$ \Ms, unequal planet masses or spacing showing the very good agreement of the analytical expression.

\section{Discussion: key concepts and generalization}

\subsection{Dynamical spacing: implication for observation and theoretical models}
\label{sec:spacing}

\begin{figure}
  \includegraphics[width=0.47\linewidth]{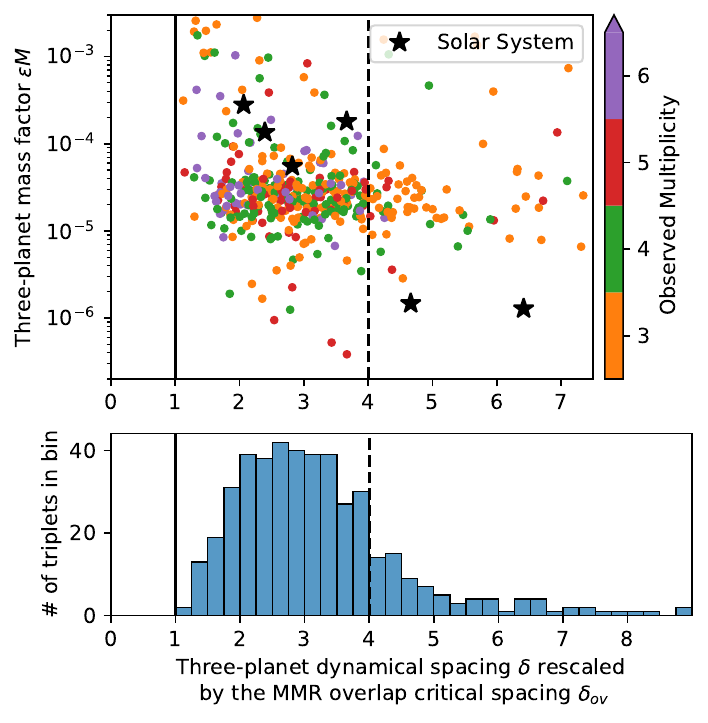}%
  \hfill%
  \includegraphics[width=0.47\linewidth]{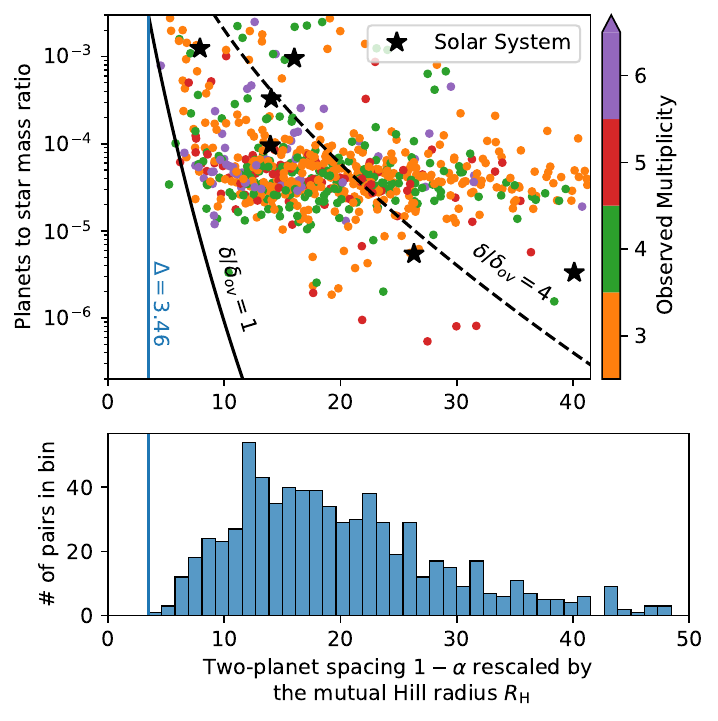}%
  \caption{Left: Three-planet dynamical spacing $\plsep$ scaled by the MMR overlap critical spacing as a function of the triplet mass factor $\epsilon M$ (top panel) and binned (bottom panel). The continuous black line is the stability condition, 80\% of the triplets have $\delta \leq 4\plsepov$ (dashed line).
  Right: Separation in mutual Hill radii for pairs of planets in 3+ planetary systems as a function of the planets-to-star mass ratio (top panel) and binned (bottom panel). The blue line at 3.46 is the Hill stability criterion, the two black lines represent the equivalent criteria from the left figure (assuming equal spacing).\label{fig:exospace}}
\end{figure}

Until now, most of the studies focused on the spacing of exoplantary systems used the Hill radius as a way to measure compactness.
However, exoplanet to star mass ratio $\epsilon$ spans from a few $10^{-7}$ to more than $10^{-3}$, such that the difference between a scaling in $\epsilon^{1/3}$ or $\epsilon^{1/4}$ is significant.
Smaller exoplanets tend to have wider spacing when expressed in Hill radii but there should be no significant mass dependency when expressed in the unit of the MMR overlap $\plsepov$ (Eq. \ref{eq:plsep-ov}).

We select from the Exoplanet Archive\footnote{\href{https://exoplanetarchive.ipac.caltech.edu/}{https://exoplanetarchive.ipac.caltech.edu/}, consulted on the 30/06/2023.} all the systems containing more than three planets to study how close to the stability limits these systems are.
For systems with only radius know, we use the mass-radius relationship from \cite{Marcy2014}. 
Due to the weak exponents at play, there is no need for a more refined relationship.
We plot on Fig. \ref{fig:exospace} the dynamical spacing $\plsep$ (Eq. \ref{eq:plsep-def}) for triplets of adjacent exoplanets rescaled by the critical value $\plsepov$ (left plot), as a histogram and as a function of the triplet mass factor $\epsilon M$.
We also plot the spacing between adjacent pairs of planets measured in Hill radii as a histogram and as a function of the planet mass ratio.
Solar System triplets and pairs are plotted as stars.
While the two mass ratio units are not the same, they allow for a meaningful comparison of the two spacing metrics.
The black continuous lines give the three-planet stability criterion for circular orbits.
On the right plot, the Hill stability limit is plotted in blue.

We see on the Hill spacing plot that a clear mass ratio dependency is visible.
As the mass ratio decreases, spacings close to the two planet limit become very rare and almost all systems are consitstent with the three planet stability criterion.
Importantly, terrestrial planets all show Hill spacing larger than 17.
On the left plot, there is no obvious mass dependency of the lower limit in spacing, terrestrial planets are consistent with the spacing value found for the bulk of the other exoplanets.

Expressed in terms of dynamical spacing $\plsep$, exoplanetary systems are much closer to the stability limit than expected from their spacing in Hill radius. 
Indeed, the Hill spacing distribution starts from the two-planet stability, have a maximum between 10 and 20 and 80\% of the pairs have a spacing lower than 30.
So the spread in spacing spans roughly an order of magnitude.
On the other hand the dynamical spacing spread is only a factor~4.

Our sample selection is subject to selection biases, for instance detected systems tend to be closer together as wide spacing leads to larger period for the outer planets \citep{Zhu2020}.
Yet, the general pattern is strong enough that the conclusions should be robust in general.
Future studies of dynamical packing should use the measure of spacing defined in this proceeding.
For pair of planets, an alternative to the Hill spacing was proposed by \cite{Tamayo2021}.

\subsection{Using the resonance optical depth}
\label{sec:optdepth}

The optical resonance depth has been shown to be an extremely powerfool tool to approach stability and diffusion problem for overlapping resonances \citep{Quillen2011,Hadden2018,Petit2020a}.
This section reviews the conditions which allow the use of this comcept.

As it relies on \cite{Chirikov1979} theory, the phase space should be shaped by a resonance network having at least some partial overlapping.
The network can be composed of very diverse resonances in terms of width, index, order, and, resonant direction.
However, an important feature is that each individual resonance can be described by a simple model when taken in isolation.
Works on generic resonances such as \cite{Hadden2019, Petit2021, Batygin2021} remain thus important to future studies on instabilities.

The resonant optical depth shines when one can exploit the symmetries of subnetworks where resonances have similar widths, spacing so that partial overlap criteria can be derived for the subnetworks.
It is then straightforward to sum over the subnetworks to obtain a general criterion.
Summing over the different subnetworks is in general possible because while resonance locii are not uniformely spread, the rational numbers are dense.
In a sense, we use to our advantage the main hurdle in the search of invariant tori that real numbers with a bad Diophantine approximation are rare.

Diffusion rates obtained under the resonant optical depth approach also have requirements for a direct application.
For a general network, each resonance defines an hypersurface in phase space and the associated chaotic diffusion takes place perpendicular to it.
As a result, the diffusion is in general anisotropic and multidimensional.
Diffusion rate can be estimated simply if the motion can be reduced to a simple direction or if one can project the random walk onto a single coordinate such that the other variables affect only weakly the diffusion rate.
In the case of compact system, this direction is $\resloc$ as the diffusion due to the three-planet MMR zeroth-order network forces the orbits to stay circular and the conservation of the total angular momentum limits the diffusion in the parallel diffusion.
In the case of a two-planet system, the conservation of angular momentum and energy binds the change in period ratio and AMD such that the motion in the $(\alpha, {\rm AMD})$ is unidimensional \citep{Petit2018}.

\subsection{More general systems: non-circular orbits, more than three planets}
\label{sec:ecc}
Stability criteria for more general systems based on three-planet MMR overlap are still an active research direction.
\cite{Tamayo2021} proposed a first step in that direction by summing the optical depths for two-planet MMR \citep{Hadden2018} for adjacent pairs of planets.
They showed that while taking into account the secular evolution was necessary, the problem was solvable.
In Bodart et al. (\emph{in prep.}) we will propose an extension of this result accounting for both the two-planet and three-planet results.

We proposed in \citep{Petit2020a} a tentative extension of the three planet result to four and more than five planet systems.
We showed that increasing the number of MMR in the network provide a heuristic result consistent with simulations.
A more in-depth study remains necessary.

\section{Conclusion}
To summarise, we estimate the instability time of compact planetary systems by modelling the diffusion rate along the zeroth-order three-planet MMR network.
We show that the complex random walk along the resonance network can be represented by a diffusion process with an effective locally constant diffusion coefficient thanks to the powerful concept of resonance optical depth.
Our estimate survival time is exponential in planet spacing as fitted in numerical simulations for the range of times of interest.
For longer times, we predict much longer survival times, which is hinted by the longest performed simulations \citep{Obertas2017}.

The mean survival timescales depends on the planet separation in units of the planet-to-star mass ratio $\epsilon^{1/4}$ and not in units of Hill radii (scaling as $\epsilon^{1/3}$).
The critical spacing for stability thus scales as $\epsilon^{1/4}$ too.
In particular, considering  systems of various masses allows both in numerical simulations and observed systems highlights the difference in scaling, proving the proposed mechanism.
Future missions for exoplanet discoveries such as Plato will increase our number of known terrestrial planets which will allow to test definitively the mechanism driving planetary system instability.

\bibliographystyle{aa}
\bibliography{cpsii.bib}

\begin{thebibliography}{32}
\expandafter\ifx\csname natexlab\endcsname\relax\def\natexlab#1{#1}\fi

\bibitem[{Batygin {et~al.}(2021)Batygin, Mardling, \& Nesvorn{\'y}}]{Batygin2021}
Batygin, K., Mardling, R.~A., \& Nesvorn{\'y}, D. 2021, The Astrophysical Journal, 920, 148

\bibitem[{Chambers {et~al.}(1996)Chambers, Wetherill, \& Boss}]{Chambers1996}
Chambers, J., Wetherill, G., \& Boss, A. 1996, Icarus, 119, 261

\bibitem[{Chirikov(1979)}]{Chirikov1979}
Chirikov, B.~V. 1979, Physics Reports, 52, 263

\bibitem[{Faber \& Quillen(2007)}]{Faber2007}
Faber, P. \& Quillen, A.~C. 2007, Monthly Notices of the Royal Astronomical Society, 382, 1823

\bibitem[{Fabrycky {et~al.}(2014)Fabrycky, Lissauer, Ragozzine, Rowe, Steffen, Agol, Barclay, Batalha, Borucki, Ciardi, Ford, Gautier, Geary, Holman, Jenkins, Li, Morehead, Morris, Shporer, Smith, Still, \& Van~Cleve}]{Fabrycky2014}
Fabrycky, D.~C., Lissauer, J.~J., Ragozzine, D., {et~al.} 2014, The Astrophysical Journal, 790, 146

\bibitem[{Gladman(1993)}]{Gladman1993}
Gladman, B. 1993, Icarus, 106, 247

\bibitem[{Hadden(2019)}]{Hadden2019}
Hadden, S. 2019, AJ, 158, 238

\bibitem[{Hadden \& Lithwick(2018)}]{Hadden2018}
Hadden, S. \& Lithwick, Y. 2018, The Astronomical Journal, 156, 95

\bibitem[{He {et~al.}(2020)He, Ford, Ragozzine, \& Carrera}]{He2020}
He, M.~Y., Ford, E.~B., Ragozzine, D., \& Carrera, D. 2020, The Astronomical Journal, 160, 276

\bibitem[{Hussain \& Tamayo(2020)}]{Hussain2020}
Hussain, N. \& Tamayo, D. 2020, Monthly Notices of the Royal Astronomical Society, 491, 5258

\bibitem[{Izidoro {et~al.}(2017)Izidoro, Ogihara, Raymond, Morbidelli, Pierens, Bitsch, Cossou, \& Hersant}]{Izidoro2017}
Izidoro, A., Ogihara, M., Raymond, S.~N., {et~al.} 2017, Monthly Notices of the Royal Astronomical Society, 470, 1750

\bibitem[{Johansen {et~al.}(2012)Johansen, Davies, Church, \& Holmelin}]{Johansen2012}
Johansen, A., Davies, M.~B., Church, R.~P., \& Holmelin, V. 2012, The Astrophysical Journal, 758, 39

\bibitem[{Laskar(1997)}]{Laskar1997}
Laskar, J. 1997, Astronomy and Astrophysics, 317, L75

\bibitem[{Laskar \& Petit(2017)}]{Laskar2017}
Laskar, J. \& Petit, A.~C. 2017, Astronomy \& Astrophysics, 605, A72

\bibitem[{Marchal \& Bozis(1982)}]{Marchal1982}
Marchal, C. \& Bozis, G. 1982, Celestial Mechanics, 26, 311

\bibitem[{Marcy {et~al.}(2014)Marcy, Isaacson, Howard, Rowe, Jenkins, Bryson, Latham, Howell, Gautier, Batalha, Rogers, Ciardi, Fischer, Gilliland, Kjeldsen, {Christensen-Dalsgaard}, Huber, Chaplin, Basu, Buchhave, Quinn, Borucki, Koch, Hunter, Caldwell, Cleve, Kolbl, Weiss, Petigura, Seager, Morton, Johnson, Ballard, Burke, Cochran, Endl, MacQueen, Everett, Lissauer, Ford, Torres, Fressin, Brown, Steffen, Charbonneau, Basri, Sasselov, Winn, {Sanchis-Ojeda}, Christiansen, Adams, Henze, Dupree, Fabrycky, Fortney, Tarter, Holman, Tenenbaum, Shporer, Lucas, Welsh, Orosz, Bedding, Campante, Davies, Elsworth, Handberg, Hekker, Karoff, Kawaler, Lund, Lundkvist, Metcalfe, Miglio, Aguirre, Stello, White, Boss, Devore, Gould, Prsa, Agol, Barclay, Coughlin, Brugamyer, Mullally, Quintana, Still, Thompson, Morrison, Twicken, D{\'e}sert, Carter, Crepp, H{\'e}brard, Santerne, Moutou, Sobeck, Hudgins, Haas, Robertson, {Lillo-Box}, \& Barrado}]{Marcy2014}
Marcy, G.~W., Isaacson, H., Howard, A.~W., {et~al.} 2014, The Astrophysical Journal Supplement Series, 210, 20

\bibitem[{Morbidelli \& Vergassola(1997)}]{Morbidelli1997}
Morbidelli, A. \& Vergassola, M. 1997, Journal of Statistical Physics, 89, 549

\bibitem[{Murchikova \& Tremaine(2020)}]{Murchikova2020}
Murchikova, L. \& Tremaine, S. 2020, The Astronomical Journal, 160, 160

\bibitem[{Obertas {et~al.}(2017)Obertas, Van~Laerhoven, \& Tamayo}]{Obertas2017}
Obertas, A., Van~Laerhoven, C., \& Tamayo, D. 2017, Icarus, 293, 52

\bibitem[{Petit(2021)}]{Petit2021}
Petit, A.~C. 2021, Celestial Mechanics and Dynamical Astronomy, 133, 39

\bibitem[{Petit {et~al.}(2017)Petit, Laskar, \& Bou{\'e}}]{Petit2017}
Petit, A.~C., Laskar, J., \& Bou{\'e}, G. 2017, Astronomy and Astrophysics, 607, A35

\bibitem[{Petit {et~al.}(2018)Petit, Laskar, \& Bou{\'e}}]{Petit2018}
Petit, A.~C., Laskar, J., \& Bou{\'e}, G. 2018, Astronomy \& Astrophysics, 617, A93

\bibitem[{Petit {et~al.}(2020)Petit, Pichierri, Davies, \& Johansen}]{Petit2020a}
Petit, A.~C., Pichierri, G., Davies, M.~B., \& Johansen, A. 2020, Astronomy and Astrophysics, 641, A176

\bibitem[{Pichierri \& Morbidelli(2020)}]{Pichierri2020}
Pichierri, G. \& Morbidelli, A. 2020, Monthly Notices of the Royal Astronomical Society, 494, 4950

\bibitem[{Pu \& Wu(2015)}]{Pu2015}
Pu, B. \& Wu, Y. 2015, The Astrophysical Journal, 807, 44

\bibitem[{Quillen(2011)}]{Quillen2011}
Quillen, A.~C. 2011, Monthly Notices of the Royal Astronomical Society, 418, 1043

\bibitem[{Smith \& Lissauer(2009)}]{Smith2009}
Smith, A.~W. \& Lissauer, J.~J. 2009, Icarus, 201, 381

\bibitem[{Tamayo {et~al.}(2021)Tamayo, Murray, Tremaine, \& Winn}]{Tamayo2021}
Tamayo, D., Murray, N., Tremaine, S., \& Winn, J. 2021, The Astronomical Journal, 162, 220

\bibitem[{Weiss {et~al.}(2018)Weiss, Marcy, Petigura, Fulton, Howard, Winn, Isaacson, Morton, Hirsch, Sinukoff, Cumming, Hebb, \& Cargile}]{Weiss2018}
Weiss, L.~M., Marcy, G.~W., Petigura, E.~A., {et~al.} 2018, The Astronomical Journal, 155, 48

\bibitem[{Wisdom(1980)}]{Wisdom1980}
Wisdom, J. 1980, The Astronomical Journal, 85, 1122

\bibitem[{Xie {et~al.}(2016)Xie, Dong, Zhu, Huber, Zheng, De~Cat, Fu, Liu, Luo, Wu, Zhang, Zhang, Zhou, Cao, Hou, Wang, \& Zhang}]{Xie2016}
Xie, J.-W., Dong, S., Zhu, Z., {et~al.} 2016, Proceedings of the National Academy of Sciences, 113, 11431

\bibitem[{Zhu(2020)}]{Zhu2020}
Zhu, W. 2020, The Astronomical Journal, 159, 188

\end{thebibliography}

\end{document}